\documentclass[cameraready]{Interspeech}
\usepackage{amsmath,graphicx}
\usepackage{booktabs} 
\usepackage{multirow} 
\usepackage{xcolor}
\usepackage{cite}

\title{Dual-LoRA: Parameter-Efficient Adversarial Disentanglement for \texorpdfstring{\\}{ } Cross-Lingual Speaker Verification}

\author[affiliation={1}]{Qituan Shangguan}{}
\author[affiliation={2,3}]{Junhao Du}{}
\author[affiliation={4,3}]{Kunyang Peng}{}
\author[affiliation={4,3}]{Feng Xue}{}
\author[affiliation={4,3}]{Hui Zhang}{}
\author[affiliation={5}]{Xinsheng Wang}{}
\author[affiliation={2,3}]{Kai Yu}{}
\author[affiliation={1}]{Shuai Wang}{}

\address{
  $^1$School of Intelligence Science and Technology, Nanjing University, Suzhou, China\\
  $^2$X-LANCE Lab, School of Computer Science, MoE Key Lab of Artificial Intelligence, Shanghai Jiao Tong University, Shanghai, China\\
  $^3$Jiangsu Key Lab of Language Computing, China\\
  $^4$AISpeech Co., Ltd., Suzhou, China; $^5$Soul AI Lab, China
}

\email{qituanshangguan@smail.nju.edu.cn, shuaiwang@nju.edu.cn}


\begin{document}

\maketitle

\begin{abstract}
Cross-lingual speaker verification suffers from severe language-speaker entanglement. This causes systematic degradation in the hardest scenario: correctly accepting utterances from the same speaker across different languages while rejecting those from different speakers sharing the same language. Standard adversarial disentanglement degrades speaker discriminability; blind discriminators inadvertently penalize speaker-discriminative traits that merely correlate with language. To address this, we propose Dual-LoRA, injecting trainable task-factorized LoRA adapters into a frozen pre-trained backbone. Our core innovation is a Language-Anchored Adversary: by grounding the discriminator with an explicit language branch, adversarial gradients target true linguistic cues rather than arbitrary correlations, preserving essential speaker characteristics. Evaluated on the TidyVoice benchmark, our system achieves a 0.91\% validation EER and achieves 3rd place in the official challenge.
\end{abstract}

\section{Introduction}
\label{sec:intro}

Speaker verification (SV) is the task of determining whether two utterances originate from the same speaker, forming the foundation of voice-based authentication and personalization systems. Large-scale pre-training has significantly advanced the field: self-supervised and foundation models such as WavLM \cite{chen2022wavlm} and w2v-BERT \cite{chung2021w2v} learn rich acoustic and phonetic representations from diverse speech. Coupled with metric learning objectives and large-margin softmax training \cite{xiang2019margin}, this pre-train-then-adapt paradigm has driven SV to remarkable maturity \cite{wang2023wespeaker,wang2024advancing}. State-of-the-art systems now achieve around $0.1$\% EER \cite{li2025enhancing} on established benchmarks like the VoxCeleb evaluation sets \cite{nagrani2017voxceleb}.

Despite this progress, performance severely degrades in complex real-world deployment scenarios \cite{baali2025sveritas}. Cross-lingual speaker verification, which verifies identity when enrollment and test utterances are in different languages, remains a critical yet underexplored challenge \cite{misra2014spoken}. Different languages introduce distinct phoneme inventories, prosodic patterns, and articulatory habits, causing high intra-speaker variability. More fundamentally, cross-lingual SV suffers from severe \textit{language-speaker entanglement} \cite{misra2014spoken}: models may exploit shared linguistic cues as proxies for speaker traits.

Prior work has approached language-speaker disentanglement primarily through adversarial training via Gradient Reversal Layers (GRL) \cite{ganin2015unsupervised}, as well as domain adaptation \cite{sun2017correlation} and factorized representation learning \cite{hsu2017unsupervised}. Specifically addressing the cross-lingual SV challenge, several targeted strategies have emerged. For instance, unsupervised adversarial domain adaptation \cite{xia2019cross} and end-to-end language adaptation frameworks \cite{rohdin2019speaker} have been utilized to align feature distributions and explicitly suppress language-specific information. Taking a slightly different perspective to mitigate language mismatch, recent work explores incorporating fine-grained phonetic information alongside speaker-sensitive feature guidance \cite{ji2025improved}. While these methods demonstrate promise, effectively disentangling language from speaker identity—without inadvertently degrading speaker discriminability—remains an ongoing challenge.

The recent TidyVoice Challenge \cite{farhadi2026tidy} provides a timely and well-defined platform to revisit this problem. By releasing a standardized cross-lingual training corpus alongside a controlled evaluation protocol, the challenge re-focuses the research community on building systems that are genuinely robust to language variation. We participated in this challenge and proposed \emph{Dual-LoRA}, which injects two parallel LoRA \cite{hu2022lora} streams into a frozen pre-trained backbone to provide dedicated parameter spaces for speaker and language information. To further prevent identity-discriminative features from being inadvertently suppressed during adversarial training, we introduce a ``Language-Anchored Adversary'' that guides the GRL discriminator using explicit language representations derived from the language LoRA stream.

Our main contributions are as follows:
\begin{itemize}
    \item We propose \textbf{Dual-LoRA}, a parameter-efficient fine-tuning framework that injects two parallel LoRA adapter streams into a frozen pre-trained backbone, enabling explicitly factorized adaptation for speaker identity and language without catastrophic forgetting.
    \item We introduce \textbf{Language-Anchored Adversarial Disentanglement}, which guides the GRL discriminator with explicit language representations, directing adversarial suppression toward linguistic cues rather than arbitrary language-correlated features.
    \item We validate the framework across multiple backbones, including ResNet \cite{he2016deep} variants and pre-trained w2v-BERT 2.0 \cite{barrault2023seamlessm4t}, observing consistent improvements across settings.
    \item On the TidyVoice benchmark, our system achieves a 2.43\% EER on the tv26\_eval-A set and a 2.84\% EER on the tv26\_eval-U set, ranking 3rd in the official evaluation.
\end{itemize}

\begin{figure*}[t]
  \centering
  \includegraphics[width=\linewidth]{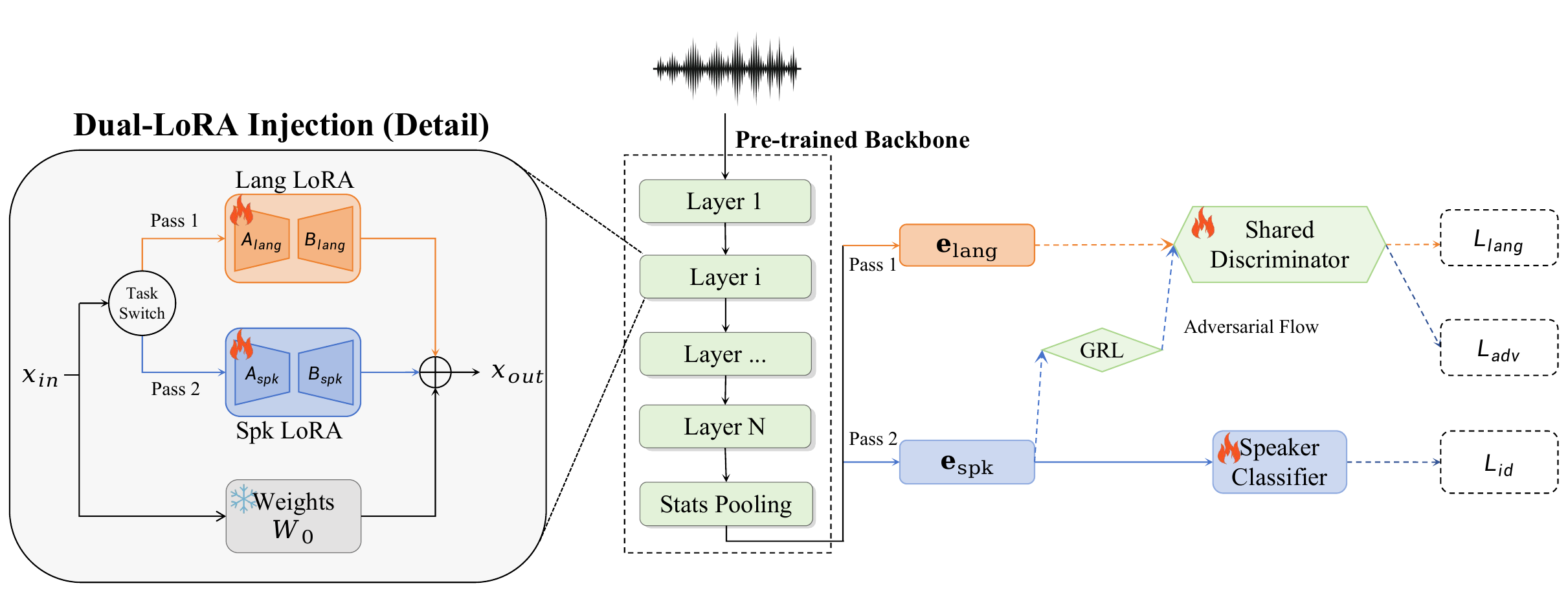}
  \caption{The overall architecture of Dual-LoRA. The framework keeps the pre-trained backbone frozen while injecting two parallel LoRA branches globally into all layers: (1) The Language Branch (top pathway, Pass 1) extracts $\mathbf{e}_{lang}$ to guide the shared discriminator; (2) The Speaker Branch (bottom pathway, Pass 2) extracts $\mathbf{e}_{spk}$. The Language-Anchored Adversarial Mechanism ensures $\mathbf{e}_{spk}$ is disentangled from linguistic content via the Gradient Reversal Layer (GRL).}
  \label{fig:architecture}
\end{figure*}

\section{Methodology}
\label{sec:method}

\subsection{Overview}
The Dual-LoRA framework addresses language-speaker entanglement in cross-lingual SV through two design principles: (1) freeze the pre-trained backbone and adapt via parallel parameter-efficient streams to preserve pre-trained generalization, and (2) guide the adversarial training by sharing a discriminator between the speaker branch and a dedicated language branch.

\subsection{The Dual-LoRA Architecture}
To effectively decouple the representation spaces of speaker identity and linguistic content without suffering from catastrophic forgetting, we construct a dual-path architecture. As illustrated in Fig.~\ref{fig:architecture}, two parallel sets of LoRA~\cite{hu2022lora} modules are injected into the frozen backbone: the Speaker Branch and the Language Branch, extracting embeddings $\mathbf{e}_{spk}$ and $\mathbf{e}_{lang}$, respectively.

\subsubsection{Disentangled LoRA Modules}
To avoid feature interference between the two branches, we allocate independent LoRA modules. For a frozen layer with weights $W_0$ and input $x$, the output is conditioned on the 
task indicator $t \in \{spk, lang\}$:
\begin{equation}
    h(x, t) = W_0 x + \Delta W_t x, \quad \Delta W_t = \frac{\alpha}{r_t} B_t A_t
\end{equation}
where $A_t \in \mathbb{R}^{r_t \times d}$ and $B_t \in \mathbb{R}^{d \times r_t}$ are 
the learnable low-rank matrices for task $t$, $r_t$ is the task-specific rank, and 
$\alpha$ is a shared scaling hyperparameter.

\subsubsection{Global Asymmetric Injection}
\label{sec:global_injection}
LoRA modules are injected globally across all backbone layers: all residual stages for ResNet~\cite{he2016deep} and all 24 layers for w2v-BERT2-Adapter-MFA~\cite{li2025enhancing}. Global injection is motivated by the observation that language-speaker entanglement spans the full feature hierarchy, from low-level phonetic to high-level prosodic representations \cite{pasad2021layer}. We further adopt an asymmetric rank design with a higher rank for the Speaker Branch ($r_{spk}=16$) and a lower rank for the Language Branch ($r_{lang}=4$), ensuring the auxiliary language branch serves as a lightweight anchor without competing with identity extraction \cite{zhang2023adalora}.

\subsection{Language-Anchored Adversarial Disentanglement}
Standard adversarial training can inadvertently compromise speaker discriminability by penalizing features where linguistic and speaker traits are inherently entangled \cite{bousmalis2016domain}. To prevent this and explicitly target genuine linguistic variations, we share a discriminator $D$ (an MLP projection head) between language classification and adversarial disentanglement.

\subsubsection{Mechanism}
Training involves two forward passes through $D$. In the Language Anchor Flow, $\mathbf{e}_{lang}$ is fed directly into $D$ for language classification, training $D$ to define decision boundaries grounded in true linguistic variation. In the Adversarial Flow, $\mathbf{e}_{spk}$ passes through a GRL before entering $D$, where the GRL reverses gradients by a factor of $-\eta$ during backpropagation \cite{ganin2015unsupervised}. Because $D$ is shared with the Anchor Flow, the adversarial gradients propagated to $\mathbf{e}_{spk}$ are linguistically grounded, targeting genuine linguistic content rather than arbitrary correlated features.

\subsubsection{Objective Functions}
The total training objective combines three losses:
\begin{equation}
    \mathcal{L}_{total} = \mathcal{L}_{id} + \lambda_{1} \mathcal{L}_{lang} + \lambda_{2} \mathcal{L}_{adv}
\end{equation}
$\mathcal{L}_{id}$ is the Sub-center ArcMargin loss~\cite{deng2020sub} applied to $\mathbf{e}_{spk}$ for speaker discrimination. $\mathcal{L}_{lang}$ and $\mathcal{L}_{adv}$ are both cross-entropy losses over $C$ language classes:
\begin{align}
    \mathcal{L}_{lang} &= -\sum_{c=1}^{C} y^{lang}_c \log [D(\mathbf{e}_{lang})]_c \\
    \mathcal{L}_{adv}  &= -\sum_{c=1}^{C} y^{lang}_c \log [D(\tilde{\mathbf{e}}_{spk})]_c
\end{align}
where $y^{lang}_c$ is the one-hot language label and 
$\tilde{\mathbf{e}}_{spk} = \text{GRL}(\mathbf{e}_{spk})$.

\subsection{Curriculum Training}
To ensure stable convergence, we adopt a three-phase curriculum~\cite{bengio2009curriculum} over three epochs. In Phase I ($\lambda_1=1.0, \lambda_2=0$), only language classification is active, allowing the language branch and discriminator to establish reliable linguistic boundaries. In Phase II ($\lambda_1=0.2, \lambda_2=0.2$), adversarial training is introduced conservatively. In Phase III ($\lambda_1=0.2, \lambda_2=0.5$), the adversarial weight is increased to enforce full disentanglement.

\subsection{Inference}
At inference, the language branch and discriminator $D$ are discarded. The speaker adaptation weights are merged into the frozen backbone via $W = W_0 + \Delta W_{spk}$ \cite{hu2022lora}, yielding zero additional computational overhead over the unadapted baseline.

\section{Experiments}
\label{sec:exp}

\subsection{Experimental Setup}
\textbf{Datasets.} We conduct evaluations on the TidyVoice Challenge dataset (TidyVoiceX)~\cite{farhadi2026tidy}, which comprises a training set (3,666 speakers, 262k utterances) and a development set (808 speakers, 60k utterances). For all single-system analyses and ablation studies (Sec.~\ref{ssec:single_sys} and \ref{ssec:ablation}), we use only public datasets (VoxBlink `VB' \cite{lin2024voxblink,lin2024voxblink2} and VoxCeleb `VC' \cite{nagrani2017voxceleb}) to ensure fair comparison. For the final challenge submission (Sec.~\ref{ssec:challenge}), we initialize the backbones using a large-scale internal multilingual corpus (approx. 18k hours, 396 languages), in compliance with the open-condition rules.

\textbf{Comparison Configurations.} We define three comparative settings to validate the Language-Anchored mechanism:
\begin{itemize}
    \item \textit{No Adv:} Dual-LoRA trained with $\lambda_2=0$, where the two branches are updated independently without gradient interaction. This serves as the primary control.
    \item \textit{Std Adv:} Standard DANN~\cite{ganin2015unsupervised} where the discriminator predicts language directly from $\mathbf{e}_{spk}$ without guidance from the language branch, representing the conventional blind disentanglement baseline.
    \item \textit{Dual-LoRA (Ours):} The proposed framework with the shared discriminator anchored by the language branch.
\end{itemize}

\textbf{Implementation Details.}
All systems are implemented in PyTorch and fine-tuned on TidyVoice for 3 epochs using the Sub-center ArcMargin loss ($K=3$) \cite{deng2020sub} with MUSAN \cite{snyder2015musan} and RIR \cite{ko2017study} augmentation. For ResNet variants (SamResNet100 \cite{yang2021simam}, ResNet293 \cite{he2016deep}), we use SGD (LR: $0.01 \to 0.0001$) with LoRA ranks $r_{spk}=16, r_{lang}=4$. For w2v-BERT2-Adapter-MFA, we use AdamW \cite{loshchilov2017decoupled} with $r_{spk}=32, r_{lang}=16$. The loss weights $(\lambda_1, \lambda_2)$ follow the three-phase curriculum described in Sec.~\ref{sec:method}.

\subsection{Single System Analysis}
\label{ssec:single_sys}
Table~\ref{tab:main_results} details the progressive improvements on the development set to establish our optimal adaptation protocol.

\textit{Impact of Loss Function (B0 vs. B1).}
Replacing standard ArcFace \cite{deng2019arcface} with Sub-center ArcMargin reduces the EER from 3.07\% to 2.05\%, confirming the necessity of multi-center modeling for the high intra-speaker variance in cross-lingual speech.

\textit{PEFT vs. Full Fine-Tuning (B1 vs. B2).}
Applying LoRA to a frozen backbone (B2) further reduces the EER to 1.66\% compared to full fine-tuning (B1). This demonstrates that PEFT effectively prevents catastrophic forgetting on limited target data.

\textit{Impact of Pre-training Data (B2 vs. B3).}
Restricting initialization data to VoxBlink (VB) yields a lower EER of 1.57\%, indicating better domain alignment with the challenge scenario than the combined VB+VC corpus.

\textit{Scalability and Disentanglement (B4 \& Ours).}
Scaling to SamResNet100 (B4) provides a strong 1.25\% baseline. Our Dual-LoRA reduces this to 0.98\%, and achieves the optimal 0.91\% EER on the w2v-BERT2 backbone, proving the effectiveness and scalability of our disentanglement strategy.

\begin{table}[t]
  \caption{Performance comparison on the TidyVoice Development Set to establish the optimal adaptation strategy. `Data' denotes pre-training data used for initialization.}
  \label{tab:main_results}
  \centering
  \footnotesize
  \setlength{\tabcolsep}{3.5pt}
  \begin{tabular}{l l l l c}
    \toprule
    ID & Backbone & Strategy & Data & EER (\%) \\
    \midrule
    B0 & SamResNet34 & Full FT (ArcFace) & VB+VC & 3.07 \\
    B1 & SamResNet34 & Full FT (Sub-center) & VB+VC & 2.05 \\
    B2 & SamResNet34 & LoRA (No Adv) & VB+VC & 1.66 \\
    B3 & SamResNet34 & LoRA (No Adv) & VB & 1.57 \\
    B4 & SamResNet100 & LoRA (No Adv) & VB & 1.25 \\
    \midrule
    Ours & SamResNet100 & Dual-LoRA & VB & 0.98 \\
    Ours & w2v-BERT2 & Dual-LoRA & VB+VC & \textbf{0.91} \\
    \bottomrule
  \end{tabular}
\end{table}

\subsection{Analysis I: Resolving the Cross-Lingual Bottleneck}
\label{ssec:worst_case}
We analyze performance across four trial scenarios defined by Same/Different Speaker (SS/DS) and Same/Different Language (SL/DL), comparing the official baseline (3.07\% overall EER) against our best Dual-LoRA system (0.91\% overall EER).

\begin{figure}[ht]
  \centering
  \includegraphics[width=0.95\linewidth]{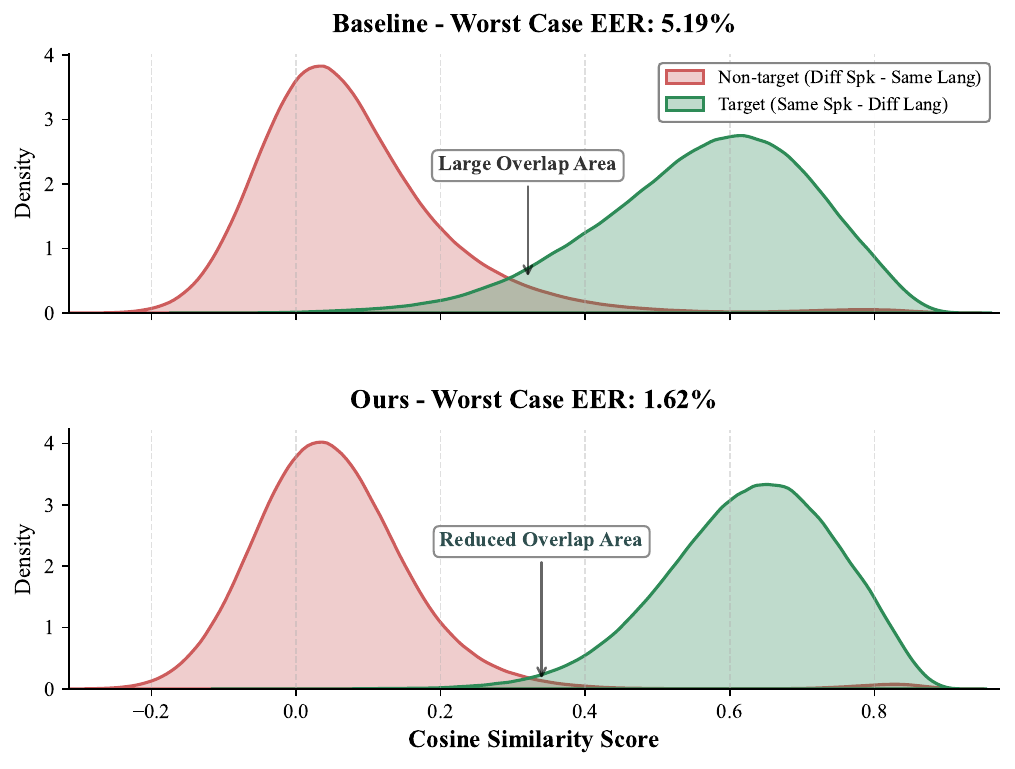} 
  \caption{Score density distribution for the worst-case scenario (SS-DL vs. DS-SL). Dual-LoRA (bottom) demonstrates significantly reduced overlap between the non-target and target distributions compared to the official baseline (top).}
  \label{fig:score_dist}
\end{figure}

As shown in Table~\ref{tab:scenario_analysis}, while our proposed framework achieves comprehensive improvements across all four conditions, the most striking gain occurs in the notoriously difficult SS-DL vs. DS-SL scenario. In this worst-case bottleneck, the system must correctly accept the same speaker speaking different languages (SS-DL) while rejecting different speakers who happen to share the same language (DS-SL). The baseline is easily misled by shared linguistic characteristics, yielding a severe 5.19\% EER. By structurally decoupling these features, Dual-LoRA drastically reduces this EER to 1.62\%. Furthermore, Fig.~\ref{fig:score_dist} shows that Dual-LoRA cleanly separates the heavily overlapping target and non-target score distributions seen in the baseline, confirming the effectiveness of our disentanglement.

\begin{table}[ht]
  \caption{EER (\%) comparison across distinct trial scenarios. (SS: Same Spk, DS: Diff Spk, SL: Same Lang, DL: Diff Lang).}
  \label{tab:scenario_analysis}
  \centering
  \footnotesize
  \setlength{\tabcolsep}{5pt}
  \begin{tabular}{l@{\quad}c@{\quad}l c c} 
    \toprule
    \multicolumn{3}{c}{\multirow{2}{*}{Scenario (Target vs. Non-target)}} & Official & Ours \\
    \multicolumn{3}{c}{} & Baseline & (Best) \\
    \midrule
    \textbf{SS-DL} & \textbf{vs.} & \textbf{DS-SL (\textit{Worst-Case})} & 5.19 & \textbf{1.62} \\
    SS-SL & vs. & DS-SL & 2.97 & 1.19 \\
    SS-DL & vs. & DS-DL & 1.79 & 0.40 \\
    SS-SL & vs. & DS-DL & 0.88 & 0.22 \\
    \midrule
    \multicolumn{3}{c}{Overall Performance} & 3.07 & 0.91 \\
    \bottomrule
  \end{tabular}
\end{table}

\subsection{Analysis II: Probing Disentanglement Effectiveness}
\label{ssec:probe_analysis}
To verify that our method effectively removes language information, we train a probe classifier \cite{alain2018understanding} to predict language labels from the frozen speaker embeddings $\mathbf{e}_{spk}$ of the w2v-BERT2 model. We report the Language Identification (LID) accuracy alongside the verification EER.

As Table~\ref{tab:probe} shows, the baseline without adversarial training (\textit{No Adv}) retains significant language information, yielding 72.71\% LID accuracy and 1.25\% EER. Standard adversarial training (\textit{Std Adv}) reduces LID accuracy to 55.03\% and EER to 0.96\%, confirming language suppression benefits verification. However, \textit{Dual-LoRA} achieves the lowest LID accuracy (49.02\%) and EER (0.91\%). This demonstrates that, compared to standard approaches, our shared discriminator removes language content more thoroughly while better preserving speaker identity.

\begin{table}[ht]
  \caption{Diagnostic probe analysis on w2v-BERT2. Lower LID Acc indicates better disentanglement; lower EER indicates better verification.}
  \label{tab:probe}
  \centering
  \footnotesize
  \setlength{\tabcolsep}{10pt}
  \begin{tabular}{l c c}
    \toprule
    Method & Probe LID Acc ($\downarrow$) & EER ($\downarrow$) \\
    \midrule
    No Adv & 72.71\% & 1.25\% \\
    Std Adv & 55.03\% & 0.96\% \\
    \textbf{Dual-LoRA} & \textbf{49.02\%} & \textbf{0.91\%} \\
    \bottomrule
  \end{tabular}
\end{table}

\subsection{Ablation Study: Generalization Across Architectures}
\label{ssec:ablation}
To verify the broad applicability of our approach, we evaluate the disentanglement strategies across various backbones, specifically extending our comparison to SamResNet100 and ResNet293. As Table~\ref{tab:ablation} shows, standard adversarial training (\textit{Std Adv}) consistently improves upon the \textit{No Adv} baseline (e.g., reducing EER from 1.25\% to 1.07\% on SamResNet100). However, \textit{Dual-LoRA} achieves the lowest EER across all models. This confirms that our language-anchored mechanism effectively generalizes across different model architectures, rather than relying on a specific network design.

\begin{table}[ht]
  \caption{Ablation study across different backbones. Pre-training data: ResNet293 (VC), SamResNet100 (VB), w2v-BERT2 (VB+VC).}
  \label{tab:ablation}
  \centering
  \footnotesize
  \setlength{\tabcolsep}{6pt}
  \begin{tabular}{l c c c}
    \toprule
    \multirow{2}{*}{Method} & \multicolumn{3}{c}{EER (\%)} \\
    \cmidrule(l){2-4}
     & ResNet293 & SamResNet100 & w2v-BERT2 \\
    \midrule
    No Adv & 1.63 & 1.25 & 1.25 \\
    Std Adv & 1.53 & 1.07 & 0.96 \\
    Dual-LoRA & \textbf{1.52} & \textbf{0.98} & \textbf{0.91} \\
    \bottomrule
  \end{tabular}
\end{table}

\subsection{Evaluation on the Official Test Set}
\label{ssec:challenge}
For the final challenge submission, we pre-train three backbones (SamResNet100, ResNet293, w2v-BERT2) on a large internal corpus and adapt them using Dual-LoRA. The system uses a 1:1:1 score-level fusion of these models after score calibration \cite{brummer2013bosaris}.

As Table~\ref{tab:eval_results} shows, the fused system achieves 2.43\% and 2.84\% EER on the \texttt{tv26\_eval-A} and \texttt{tv26\_eval-U} sets, respectively. This yields over 70\% relative error reduction against the baseline, securing 3rd place. Furthermore, consistent performance across seen and unseen languages demonstrates our strategy learns generic representations without overfitting to specific patterns.

\begin{table}[ht]
  \caption{Final results on the official TidyVoice test set. `Internal' indicates large-scale multilingual initialization.}
  \label{tab:eval_results}
  \centering
  \footnotesize
  \setlength{\tabcolsep}{6pt}
  \begin{tabular}{l c c c c}
    \toprule
    \multirow{2}{*}{System} & Pre-train & Dev Set & \multicolumn{2}{c}{Eval Set EER (\%)} \\
    \cmidrule(lr){4-5}
     & Data & EER (\%) & eval-A & eval-U \\
    \midrule
    Official Baseline & VB+VC & 3.07 & 9.06 & 11.59 \\
    \midrule
    Ours (Fusion) & \textit{Internal} & 0.73 & \textbf{2.43} & \textbf{2.84} \\
    \bottomrule
  \end{tabular}
\end{table}

\section{Conclusion}
\label{sec:conclusion}

We address severe language-speaker entanglement in cross-lingual speaker verification by proposing Dual-LoRA. This parameter-efficient framework adapts frozen backbones using parallel LoRA streams to separately capture speaker and language information. To prevent the unintended identity loss in standard adversarial training, we introduce a Language-Anchored Adversary. By sharing a discriminator between an explicit language classification task and the adversarial flow, this mechanism ensures that adversarial gradients precisely target genuine linguistic variations. 

Evaluations on the TidyVoice benchmark confirm our approach's effectiveness. Dual-LoRA resolves the worst-case cross-lingual bottleneck (SS-DL vs. DS-SL), reducing EER from 5.19\% to 1.62\%. Probing demonstrates our method removes language content more thoroughly than standard baselines while preserving core identity. Ultimately, our fused system secured 3rd place on official evaluation sets, proving strong generalization across seen and unseen languages.

\section{Generative AI Use Disclosure}
Large language models were used only for language polishing and grammatical correction. All scientific content, experimental design, and data analysis are the original work of the authors.

\bibliographystyle{IEEEtran}
\bibliography{mybib} 

@article{chen2022wavlm,
  title={{WavLM}: Large-scale self-supervised pre-training for full stack speech processing},
  author={Chen, Sanyuan and Wang, Chengyi and Chen, Zhengyang and Wu, Yu and Liu, Shujie and Chen, Zhuo and Li, Jinyu and Kanda, Naoyuki and Yoshioka, Takuya and Xiao, Xiong and others},
  journal={IEEE Journal of Selected Topics in Signal Processing},
  volume={16},
  number={6},
  pages={1505--1518},
  year={2022},
  publisher={IEEE}
}

@inproceedings{chung2021w2v,
  title={{W2V-BERT}: Combining contrastive learning and masked language modeling for self-supervised speech pre-training},
  author={Chung, Yu-An and Zhang, Yu and Han, Wei and Chiu, Chung-Cheng and Qin, James and Pang, Ruoming and Wu, Yonghui},
  booktitle={2021 IEEE Automatic Speech Recognition and Understanding Workshop (ASRU)},
  pages={244--250},
  year={2021},
  organization={IEEE}
}

@misc{farhadi2026tidy,
      title={{TidyVoice}: A Curated Multilingual Dataset for Speaker Verification Derived from {Common Voice}}, 
      author={Aref Farhadipour and Jan Marquenie and Srikanth Madikeri and Eleanor Chodroff},
      year={2026},
      journal={ICASSP2026},
      url={https://arxiv.org/abs/2601.16358}, 
}

@inproceedings{misra2014spoken,
  title={Spoken language mismatch in speaker verification: An investigation with nist-sre and crss bi-ling corpora},
  author={Misra, Abhinav and Hansen, John HL},
  booktitle={2014 IEEE spoken language technology workshop (SLT)},
  pages={372--377},
  year={2014},
  organization={IEEE}
}

@article{hu2022lora,
  title={{LoRA}: Low-rank adaptation of large language models.},
  author={Hu, Edward J and Shen, Yelong and Wallis, Phillip and Allen-Zhu, Zeyuan and Li, Yuanzhi and Wang, Shean and Wang, Liang and Chen, Weizhu and others},
  journal={International Conference on Learning Representations},
  year={2022}
}

@inproceedings{ganin2015unsupervised,
  title={Unsupervised domain adaptation by backpropagation},
  author={Ganin, Yaroslav and Lempitsky, Victor},
  booktitle={International conference on machine learning},
  pages={1180--1189},
  year={2015},
  organization={PMLR}
}

@article{li2025enhancing,
  title={Enhancing Speaker Verification with w2v-BERT 2.0 and Knowledge Distillation guided Structured Pruning},
  author={Li, Ze and Cheng, Ming and Li, Ming},
  journal={arXiv preprint arXiv:2510.04213},
  year={2025}
}

@inproceedings{bengio2009curriculum,
  title={Curriculum learning},
  author={Bengio, Yoshua and Louradour, J{\'e}r{\^o}me and Collobert, Ronan and Weston, Jason},
  booktitle={Proceedings of the 26th annual international conference on machine learning},
  pages={41--48},
  year={2009}
}

@inproceedings{he2016deep,
  title={Deep residual learning for image recognition},
  author={He, Kaiming and Zhang, Xiangyu and Ren, Shaoqing and Sun, Jian},
  booktitle={Proceedings of the IEEE conference on computer vision and pattern recognition},
  pages={770--778},
  year={2016}
}

@inproceedings{deng2020sub,
  title={Sub-center arcface: Boosting face recognition by large-scale noisy web faces},
  author={Deng, Jiankang and Guo, Jia and Liu, Tongliang and Gong, Mingming and Zafeiriou, Stefanos},
  booktitle={European Conference on Computer Vision},
  pages={741--757},
  year={2020},
  organization={Springer}
}

@inproceedings{wang2023wespeaker,
  title={{WeSpeaker}: A research and production oriented speaker embedding learning toolkit},
  author={Wang, Hongji and Liang, Chengdong and Wang, Shuai and Chen, Zhengyang and Zhang, Binbin and Xiang, Xu and Deng, Yanlei and Qian, Yanmin},
  booktitle={ICASSP 2023-2023 IEEE International Conference on Acoustics, Speech and Signal Processing (ICASSP)},
  pages={1--5},
  year={2023},
  organization={IEEE}
}

@article{wang2024advancing,
  title={Advancing speaker embedding learning: Wespeaker toolkit for research and production},
  author={Wang, Shuai and Chen, Zhengyang and Han, Bing and Wang, Hongji and Liang, Chengdong and Zhang, Binbin and Xiang, Xu and Ding, Wen and Rohdin, Johan and Silnova, Anna and others},
  journal={Speech Communication},
  volume={162},
  pages={103104},
  year={2024},
  publisher={Elsevier}
}

@inproceedings{lin2024voxblink,
  title={{VoxBlink}: A large scale speaker verification dataset on camera},
  author={Lin, Yuke and Qin, Xiaoyi and Zhao, Guoqing and Cheng, Ming and Jiang, Ning and Wu, Haiying and Li, Ming},
  booktitle={ICASSP 2024-2024 IEEE International Conference on Acoustics, Speech and Signal Processing (ICASSP)},
  pages={10271--10275},
  year={2024},
  organization={IEEE}
}

@article{nagrani2017voxceleb,
  title={Voxceleb: a large-scale speaker identification dataset},
  author={Nagrani, Arsha and Chung, Joon Son and Zisserman, Andrew},
  journal={arXiv preprint arXiv:1706.08612},
  year={2017}
}

@article{snyder2015musan,
  title={{MUSAN}: A music, speech, and noise corpus},
  author={Snyder, David and Chen, Guoguo and Povey, Daniel},
  journal={arXiv preprint arXiv:1510.08484},
  year={2015}
}

@inproceedings{ko2017study,
  title={A study on data augmentation of reverberant speech for robust speech recognition},
  author={Ko, Tom and Peddinti, Vijayaditya and Povey, Daniel and Seltzer, Michael L and Khudanpur, Sanjeev},
  booktitle={2017 IEEE international conference on acoustics, speech and signal processing (ICASSP)},
  pages={5220--5224},
  year={2017},
  organization={IEEE}
}

@article{loshchilov2017decoupled,
  title={Decoupled weight decay regularization},
  author={Loshchilov, Ilya and Hutter, Frank},
  journal={arXiv preprint arXiv:1711.05101},
  year={2017}
}

@inproceedings{yang2021simam,
  title={Simam: A simple, parameter-free attention module for convolutional neural networks},
  author={Yang, Lingxiao and Zhang, Ru-Yuan and Li, Lida and Xie, Xiaohua},
  booktitle={International conference on machine learning},
  pages={11863--11874},
  year={2021},
  organization={PMLR}
}

@inproceedings{deng2019arcface,
  title={Arcface: Additive angular margin loss for deep face recognition},
  author={Deng, Jiankang and Guo, Jia and Xue, Niannan and Zafeiriou, Stefanos},
  booktitle={Proceedings of the IEEE/CVF conference on computer vision and pattern recognition},
  pages={4690--4699},
  year={2019}
}

@article{brummer2013bosaris,
  title={The bosaris toolkit: Theory, algorithms and code for surviving the new dcf},
  author={Br{\"u}mmer, Niko and De Villiers, Edward},
  journal={arXiv preprint arXiv:1304.2865},
  year={2013}
}

@article{lin2024voxblink2,
  title={{VoxBlink2}: A 100k+ speaker recognition corpus and the open-set speaker-identification benchmark},
  author={Lin, Yuke and Cheng, Ming and Zhang, Fulin and Gao, Yingying and Zhang, Shilei and Li, Ming},
  journal={arXiv preprint arXiv:2407.11510},
  year={2024}
}

@inproceedings{xia2019cross,
  title={Cross-lingual text-independent speaker verification using unsupervised adversarial discriminative domain adaptation},
  author={Xia, Wei and Huang, Jing and Hansen, John HL},
  booktitle={ICASSP 2019-2019 IEEE International Conference on Acoustics, Speech and Signal Processing (ICASSP)},
  pages={5816--5820},
  year={2019},
  organization={IEEE}
}

@inproceedings{rohdin2019speaker,
  title={Speaker verification using end-to-end adversarial language adaptation},
  author={Rohdin, Johan and Stafylakis, Themos and Silnova, Anna and Zeinali, Hossein and Burget, Luk{\'a}{\v{s}} and Plchot, Old{\v{r}}ich},
  booktitle={ICASSP 2019-2019 IEEE International Conference on Acoustics, Speech and Signal Processing (ICASSP)},
  pages={6006--6010},
  year={2019},
  organization={IEEE}
}

@inproceedings{ji2025improved,
  title={Improved Cross-Lingual Speaker Verification Using Speaker Sensitive Feature Guidance and Fine-grained Phonetic Information},
  author={Ji, Yongtai and Li, Guangxing and Huang, Hao and Li, Yanbing and Silamu, Wushour},
  booktitle={ICASSP 2025-2025 IEEE International Conference on Acoustics, Speech and Signal Processing (ICASSP)},
  pages={1--5},
  year={2025},
  organization={IEEE}
}

@inproceedings{baali2025sveritas,
  title={SVeritas: Benchmark for Robust Speaker Verification under Diverse Conditions},
  author={Baali, Massa and Bisht, Sarthak and Teixeira, Francisco and Shapovalenko, Kateryna and Singh, Rita and Raj, Bhiksha},
  booktitle={Findings of the Association for Computational Linguistics: EMNLP 2025},
  pages={9714--9731},
  year={2025}
}

@inproceedings{xiang2019margin,
  title={Margin matters: Towards more discriminative deep neural network embeddings for speaker recognition},
  author={Xiang, Xu and Wang, Shuai and Huang, Houjun and Qian, Yanmin and Yu, Kai},
  booktitle={2019 Asia-Pacific Signal and Information Processing Association Annual Summit and Conference (APSIPA ASC)},
  pages={1652--1656},
  year={2019},
  organization={IEEE}
}

@incollection{sun2017correlation,
  title={Correlation alignment for unsupervised domain adaptation},
  author={Sun, Baochen and Feng, Jiashi and Saenko, Kate},
  booktitle={Domain adaptation in computer vision applications},
  pages={153--171},
  year={2017},
  publisher={Springer}
}

@article{hsu2017unsupervised,
  title={Unsupervised learning of disentangled and interpretable representations from sequential data},
  author={Hsu, Wei-Ning and Zhang, Yu and Glass, James},
  journal={Advances in neural information processing systems},
  volume={30},
  year={2017}
}

@article{barrault2023seamlessm4t,
  title={Seamlessm4t: Massively multilingual \& multimodal machine translation},
  author={Barrault, Lo{\"\i}c and Chung, Yu-An and Meglioli, Mariano Cora and Dale, David and Dong, Ning and Duquenne, Paul-Ambroise and Elsahar, Hady and Gong, Hongyu and Heffernan, Kevin and Hoffman, John and others},
  journal={arXiv preprint arXiv:2308.11596},
  year={2023}
}

@article{bousmalis2016domain,
  title={Domain separation networks},
  author={Bousmalis, Konstantinos and Trigeorgis, George and Silberman, Nathan and Krishnan, Dilip and Erhan, Dumitru},
  journal={Advances in neural information processing systems},
  volume={29},
  year={2016}
}

@inproceedings{pasad2021layer,
  title={Layer-wise analysis of a self-supervised speech representation model},
  author={Pasad, Ankita and Chou, Ju-Chieh and Livescu, Karen},
  booktitle={2021 IEEE Automatic Speech Recognition and Understanding Workshop (ASRU)},
  pages={914--921},
  year={2021},
  organization={IEEE}
}

@article{alain2018understanding,
  title={Understanding intermediate layers using linear classifier probes, 2018},
  author={Alain, Guillaume and Bengio, Yoshua},
  journal={URL https://arxiv. org/abs/1610.01644},
  volume={1610},
  year={2018}
}

@article{zhang2023adalora,
  title={Adalora: Adaptive budget allocation for parameter-efficient fine-tuning},
  author={Zhang, Qingru and Chen, Minshuo and Bukharin, Alexander and Karampatziakis, Nikos and He, Pengcheng and Cheng, Yu and Chen, Weizhu and Zhao, Tuo},
  journal={arXiv preprint arXiv:2303.10512},
  year={2023}
}

\end{document}